# Robust acoustic cloaking with density-near-zero materials

*Jiajun Zhao[1,2], Tiancheng Han[1], Zhining Chen[1], Baowen Li[2,3], and Cheng-Wei Qiu[1,\*]*

*1 Department of Electrical and Computer Engineering, National University of Singapore, Singapore 117576, Republic of Singapore*

*2 Department of Physics and Centre for Computational Science and Engineering, National University of Singapore, Singapore 117546, Republic of Singapore*

*3 Center for Phononics and Thermal Energy Science, School of Physics Science and Engineering, Tongji University, Shanghai 200092, China*

**Abstract.**

Using the phenomenon of extraordinary sound transmission (EST) through artificially constructed density-near-zero (DNZ) materials, we propose an alternative approach to realize a very robust acoustic cloaking, which includes one absorber and one projector connected by an energy channel. The elementary unit cell is made of one single-piece homogeneous copper, but it well maintains both the original planar wavefront and the nearly perfect one-dimensional transmission, in the presence of any inserted objects. Furthermore, the overall cloaked space can be increased, designed, and distributed by arbitrarily assembling additional DNZ cells without any upper limit in the total cloaking volume. We demonstrate acoustic cloaking by different combinations of DNZ cells in free space as well as in acoustic waveguides, which enables any objects inside the cells imperceptible along undistorted sound paths.

**PACS:**

The research on invisibility cloaking has been a hotspot over decades. One of the well-known realization schemes proposed by Pendry [1] and Leohardt [2] is coordinate transformation, which requires exotic material parameters (inhomogeneous, anisotropic) [3,4]. Years before, transformation scheme of acoustic waves was theoretically proposed [5], and the experimental realization of an acoustic cloak was reported using metamaterials [6,7]. Nevertheless, the anisotropic and spatially-varying acoustic inertia or modulus based on coordinate transformation usually cannot be obtained directly from nature, resulting in complexity in realization. To overcome this difficulty, another scheme named the inversed design is adopted by García-Chocano [8] and Sanchis [9] to topologically cancel acoustic scattering caused by a given central object. It only requires a certain computationally-optimized distribution of rigid boundaries around the hidden object, without the consideration of any anisotropic or inhomogeneous parameter. However, the computationally-optimized patterns of topological acoustic cloaks highly rely on the shape and the location of the objects to be hidden, which may need to re-design for another object given.

In order to combine the advantages of the current two methods together, i.e., making an acoustic cloak independent of the objects to be hidden and simultaneously only including uniform parameters, it is essential to propose another scheme with a distinct mechanism. Recently, the combination of lumped short tubes and transverse membranes is investigated to comprise the effective DNZ material which enables EST [10]. Later, this research is extended to the realization of DNZ ultra-narrow waveguides which empower energy squeezing [11]. Currently, EST is only attainable in waveguides other than in free space. Once any object is placed inside the waveguides, EST will be totally ruined by the breakdown of the systematic resonances. However, it seems that we can adopt EST caused by the DNZ materials with some modification as a potential alternative to realize a uniform acoustic cloak in free space as well as in waveguides. Then, the crucial challenge is how to conceal the objects along sound paths in free space, while maintain the perfect plane-wave transmission independently of the hidden objects.

As illustrated in Fig. 1(a), an inhomogeneous acoustic cloak based on coordinate transformation [12] renders an object invisible by distorting its ambient flow. Here, we propose the uniform DNZ acoustic-cloaking cell as sketched in Fig. 1(b) that allows EST, hides arbitrary inserted objects, and preserves wavefronts as well as phases. Schematically, the flow at the front of an object is concentrated into the energy channel through an absorber. After, the energy is coupled out by a projector to the back where the flow is restored. In this scheme, acoustic cloaking for one-dimensional invisibility along sound paths through uniform materials is achievable without wave distortion. More importantly, our structure is irrelevant to the positions and the shapes of hidden objects, different from the inverse design [9]. The process resembles the engineering optical camouflage: positioning cameras upon an object wrapped by a retro-reflecting coat; taking pictures and transmitting the signal; projecting the front scene onto the back of the coat. Thus the object is visually concealed [13].

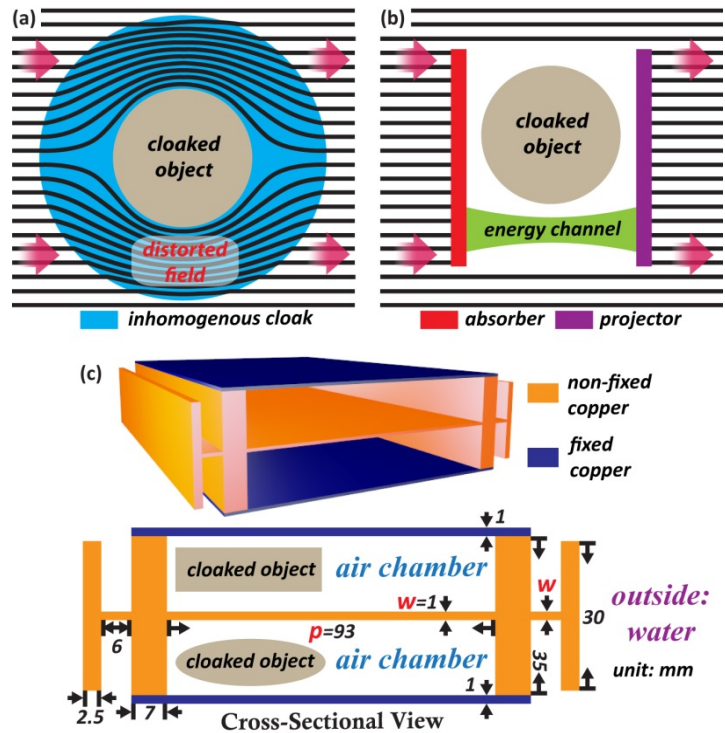

**Figure 1** (a) Acoustic cloaking based on coordinate transformation. (b) The DNZ acoustic-cloaking cell, without perturbation of the ambient flow. (c) Schematic of the DNZ cell immersed in water, which only consists of copper with air chambers inside.

For the detailed description of a DNZ acoustic-cloaking cell in Fig. 1(c), it is only implemented by copper (density: $8900 kg/m^3$; Young's modulus: 122GPa; Poisson's ratio: 0.35) [14] with two copper planks fixed without displacement. The parameters are labeled while $p$ and $w$ are the length and the width of the energy channel respectively. The two hollow enclosed chambers are filled with air (density: $1.21 kg/m^3$; speed of sound: $343 m/s$), inside which objects can be placed. Outside is water (density: $998 kg/m^3$; speed of sound: $1481 m/s$) and a monochromatic acoustic plane wave propagates from left.

The rationale of the structure is the analog of a lumped short tube containing a transverse membrane [11]. In Ref. [11], the lumped tube can only provide acoustic inertance while the massless membrane only offers acoustic compliance [15]. Therefore, only by their combination can the resonance be produced at certain frequencies. When the resonance occurs, the dynamic mass of the waveguides will effectively equal zero, leading to EST [11]. Here instead, we adopt the structure of vibrant copper immersed in water, which has acoustic inertance because of its mass and simultaneously has acoustic compliance because of its elasticity [14]. Consequently, the DNZ cell immersed in water itself constitutes a resonant device, allowing EST. Moreover, in order to isolate any objects inside the air chambers from systematic resonances, the extreme acoustic impedance mismatch between the air inside and the copper structure is utilized. Thus, the resonances are fixed regardless of any objects inside the air chambers, and the input energy will be totally transmitted as EST at the resonances, making acoustic cloaking by DNZ materials possible.

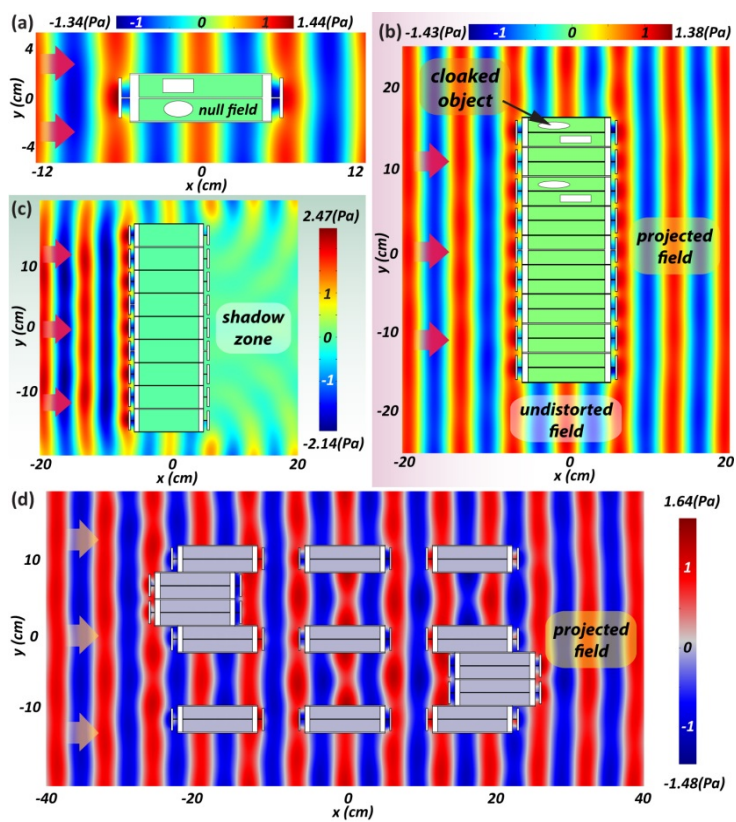

**Figure 2** Acoustic plane waves propagate in water from left in free space. (a) The two objects are hidden. There is no sound inside air chambers and the field outside the cell is almost undisturbed. (b) The DNZ array is immersed in water, at whose back the wavefront and the phase are restored. (c) When the energy channels are removed, strong scattering occurs. (d) The combination of DNZ cells forms an S-shaped cloaked space.

In order to exhibit acoustic cloaking, we first test the DNZ cell when plane waves with unit magnitude and frequency 23.1kHz propagate in water from left in Fig. 2(a). The energy channel vibrates longitudinally, and inside the air chambers there is no sound as expected, meaning the inserted objects are isolated and will not disturb the resonance and the field outside. It is noteworthy that the transmission of the input sound is nearly perfect, implying EST through the cell without backscattering. In addition, the planar wavefront and the continuous phase are restored after EST. Moreover, thanks to the two fixed planks in Fig. 1(c), we can connect many cells sharing the planks, forming an arbitrarily-designed cloaked space in free space. In Fig. 2(b), the cells are aligned in water to increase the acoustic cloaking volume. The effect is demonstrated, as if acoustic plane waves keep propagating without awareness of the bulky array as well as the multiple macroscopic objects inside. If the energy channels are removed, strong backscattering occurs in Fig. 2(c) because the resonance is ruined. We can also design the overall cloaked space with an arbitrary distribution of the cells in free space, as shown in Fig. 2(d) where an S-shaped cloaked space is formed. Note that based on DNZ cells, we accomplish acoustic cloaking in free space only by a single kind of uniform material in a feasible structure.

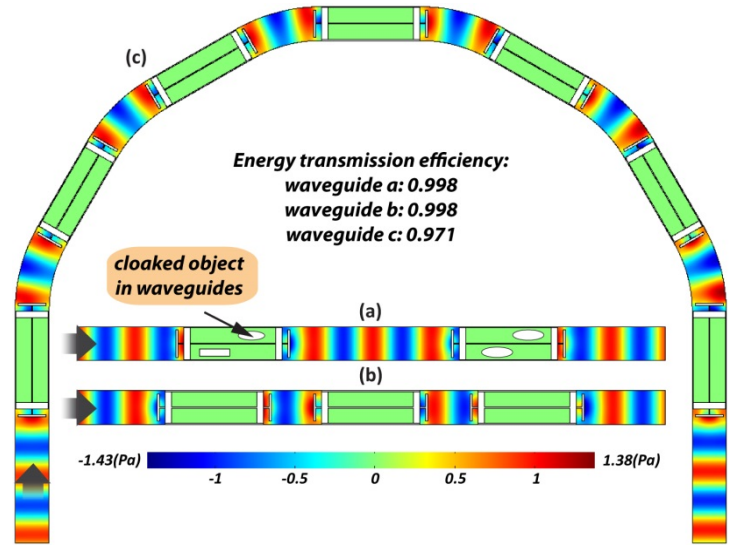

**Figure 3** Acoustic plane waves propagate in water in waveguides. (a) The four objects are hidden. There is no sound inside the air chambers and the sound outside cells is totally transmitted. (b) The resonant frequency of EST is irrelevant to the number of DNZ cells. (c) The cells can be used to hide cloaked objects as well as to bend sounds.

Apart from the scenario in free space, sound manipulation in waveguides has shown significant breakthrough, such as the acoustic circulator based on nonreciprocity without relying on nonlinearity of acoustic waves [16]. Here, the DNZ cell is functional in cloaking objects not only in free space but also in waveguides. As shown in Fig. 3(a), when acoustic waves with unit magnitude and the same frequency 23.1kHz propagate in water in hard-wall waveguides from left, the objects inside the four air chambers will be unperceivable. Note that one typical feature of DNZ materials is that energy tunneling occurs independently of the number of DNZ segments because each DNZ segment will resonance at the same frequency without the influence from others [11]. When we add more cells inside the waveguide, the resonant frequency 23.1kHz will not shift as proven in Fig. 3(b), becoming the evidence of the DNZ property of the cells. The energy transmission efficiency is almost 100% in both cases, indicating EST through DNZ cells. In other words, EST in our cases is not due to other effects such as Fabry-Pérot resonance which highly depends on the length of waveguides [17]. It is also noteworthy that by our layout the objects are cloaked along through the sound path. This is not realizable if we simply consider the insertion of narrow waveguides, which redirect the sound path to the narrow branches away from the objects [11]. In Fig. 3(c), we demonstrate the versatility of DNZ cells, which are used to bend sounds in waveguides as well as to maintain the cloaking effect.

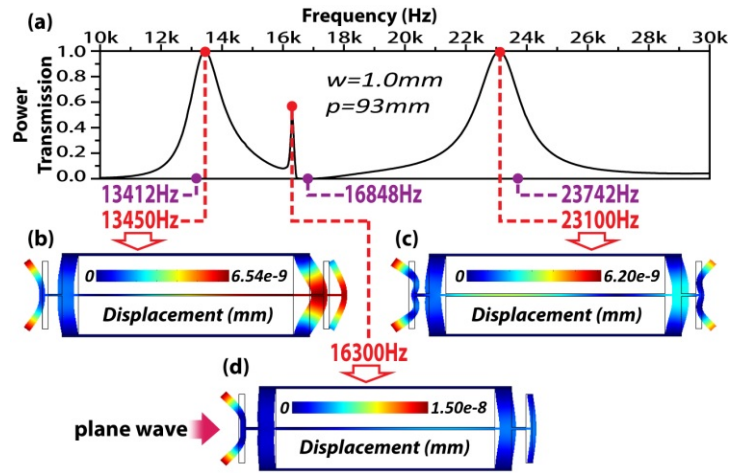

**Figure 4** (a) The power transmission at normal incidence versus the input frequency. The frequencies for the three peaks are indicated as the red dots while the eigenfrequencies of the copper parts are indicated as the purple dots. (b,c,d) The total displacements of one DNZ cell at 13450Hz, 16.3kHz and 23.1kHz.

The power transmission through DNZ acoustic-cloaking cells in Fig. 2 and Fig. 3 is presented in Fig. 4(a). There are two frequencies allowing EST (13450Hz and 23.1kHz) with a small peak (16.3kHz) in between, between which 23.1kHz is the one we adopt for the incidence in both Fig. 2 and Fig. 3. We also calculate the vibration statuses at the three frequencies in Fig. 3(b,c,d). It is noticed that the resonance-like spectrum confirms that EST only occurs at the single frequencies where the combination of acoustic inertance and acoustic compliance leads to zero density [11]. Fig. 3(b,c) imply two different resonant modes of the DNZ cell immersed in water. Further, we investigate the eigenfrequencies of the copper parts, whose values are marked as the purple dots in Fig. 3(a). The similarity between the frequencies corresponding to the red and the purple dots also verifies that cloaking lies in systematic resonances. Here, the discrepancy between the dots of different colors is because the acoustic load of the ambient water is within consideration in Fig. 3(b,c) while for the purple dots we simply consider the eigenfrequencies of the copper parts. However, the exploration of the eigenfrequencies will benefit the rough estimation of the resonant-like spectrum.

The resonant-like spectrum inspires us to understand the DNZ property of the cell again by means of a spring-mass model which is the typical mechanical translation of classical acoustic systems [18]. The cell immersed in water in Fig. 5(a) is treated here as the spring-mass model in Fig. 5(b). In Fig. 5(a), different parts of the cell are marked with various colors to illustrate the corresponding elements in Fig. 5(b). In detail, the fixed copper planks (black) are reasonable to be considered as the fixed wall (black); the input sound and the driving force are the physical counterparts (pink); so are the main body of the cell and the mass chunk (blue). The joints (red) as well as the elasticity of the cell are rationally interpreted as the ideal spring (red). We can then establish the equation for this spring-mass model:

$$M\, d^2x/dt^2 = -kx + F_0 \cos \omega t, \tag{1}$$

where $k$ is the stiffness of the ideal spring in Fig. 5(b), $M$ the mass of the chunk, $x$ the displacement, $\omega$ the driving frequency and $F_0 \cos \omega t$ the driving force. The solution of Eq. (1) is $x = A\cos \omega t$, where $A = F_0 / \left( M \left| \omega_0^2 - \omega^2 \right| \right)$. The maximum amplitude occurs at the resonance $\omega_0 = \sqrt{k/M}$. For the counterpart in Fig. 5(a), the force from input waves drives the vibration of the DNZ cell. When the resonance occurs, the energy from the input will be accumulated through the vibration system and then conveyed to its acoustic loads, i.e., the ambient media. In other words, the momentum gain of the acoustic loads, i.e., the air inside and the water outside, consumes all input power at systematic resonances.

Moreover, since the acoustic impedance of the water outside is extremely higher than that of the air inside (3561 times), the acoustic load of this system is almost completely attributed to water, affecting the resonant frequencies. Consequently, all input power is transferred to the water outside at resonances, leading to EST. In general, the air chambers not only serve as the cloaked space, but also decouple systematic resonances from the influence of the objects inside them, making the DNZ property of the cell at resonances not affected by the existence of the hidden objects. In addition, instead of the single resonance in Fig. 5(b), there must be several and various resonances of the DNZ cell in water due to the rich oscillation modes of a solid. The spring-mass model here is simply one of the comprehensive interpretations to expound the mechanism of EST and acoustic cloaking.

More importantly, the spring-mass model also implies the DNZ property of the cell at resonances, which is the acoustic equivalence of an electromagnetic epsilon-near-zero material [19]. We can define the effective mass of the vibration system in Fig. 5(a,b) as $M_{eff}(\omega) = M - k/\omega^2$, which already intrinsically includes the acoustic inertance caused by its mass as well as the acoustic compliance caused by its elasticity. Then, we can arrange Eq. (1) if the harmonic vibration is considered [10]:

$$F_0 \cos \omega t = M_{eff}(\omega) d^2x/dt^2, \qquad (2)$$

which turns into Newton's second law $force = mass \times d^2x/dt^2$. At the systematic resonance $\omega_0 = \sqrt{k/M}$, $M_{eff}(\omega) = M - k/\omega^2$ becomes zero, by which the power transmission of input sounds will be extremely enhanced [10]. Herein, the DNZ property of the cell at resonances is again explained by the aid of the spring-mass model in Fig. 5(a,b).

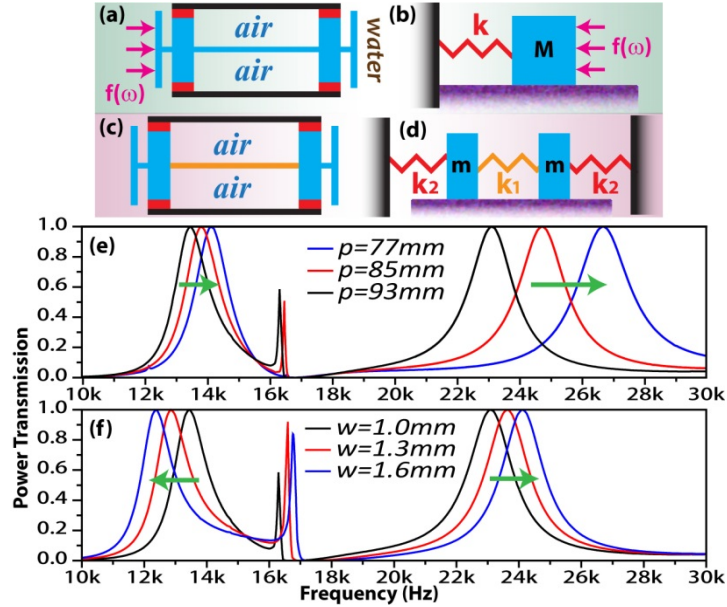

**Figure 5** (a)(c) Schematics of the DNZ acoustic-cloaking cell. Different colors stand for different portions of the cell. (b) The spring-mass model. (d) The coupling model. (e)(f) The power transmission versus the input frequency in Fig. 2 and Fig. 3 when $p$ or $w$ of the energy channel changes.

After exploring the mechanism of the DNZ acoustic-cloaking cell, we further examine the effect of the energy-channel geometry upon the spectrum by employing the coupling model in Fig. 5(c,d). Analogously to the discussion about Fig. 5(a,b), the two fixed copper planks in Fig. 5(c) (black) can be referred as the two hard walls (black) in Fig. 5(d). In order to investigate the relation between the structural geometry and the resonant frequencies, the main bodies in Fig. 5(c) are modeled individually. Then, the two copper bodies (blue) are the counterparts of the two chunks (blue) with mass $m$. The four joints (red) indicating elasticity resemble the two ideal springs (red) with stiffness $k_2$, connecting the chunks to the walls. Additionally, owing to the narrowness of the energy channel (orange), we can model it as the spring with stiffness $k_1$ and mass $m_s$ (orange), coupling the two chunks. Via the coupling model, we manage to unveil the relation of the structural geometry and the resonances of the DNZ cell.

We establish the equations for the coupling model in Fig. 5(c,d):

$$\begin{cases} m\,d^2x_1/dt^2 = -k_2 x_1 + k_1(x_2 - x_1) \\ m\,d^2x_2/dt^2 = -k_1(x_2 - x_1) - k_2 x_2 \end{cases}, \tag{3}$$

where $x_1$ and $x_2$ are the displacements of the left and the right chunks. The resonant frequencies of Eq. (3) are the lower $\omega_L = \sqrt{k_2/(m+m_s)}$ and the higher $\omega_H = \sqrt{(2k_1+k_2)/(m+m_s)}$. If the energy-channel length $p=93mm$ in Fig. 1(c) becomes longer, $m_s$ will grow larger but $k_1$ will get smaller, similar to the series connection of springs. Therefore, both $\omega_L$ and $\omega_H$ will become smaller, implying all resonant frequencies of the DNZ cell will be shifted lower when $p$ becomes longer. Contrarily, if $p$ is shortened the resonant frequencies will be shifted higher. The shift in Fig. 5(e) verifies the analysis (moving from the black curve to the red, then to the blue). On the other hand, if the energy-channel width $w=1.0mm$ becomes thicker, $m_s$ will grow larger as well as $k_1$, similar to the shunt connection of springs. Hence, the lower resonance $\omega_L$ will become lower. As for $\omega_H$, the double increments of $k_1$ at the numerator $(2k_1+k_2)$ is intuitively larger than the increment of $m_s$ at the denominator, leading to the rise of $\omega_H$. Thus, it is predicted if $w$ gets thicker, the lower resonant frequencies of the DNZ cell will be shifted even lower, whereas the higher will be even higher. The broadening of the spectrums in Fig. 5(f) verifies our analysis (moving from the black curve to the red, then to the blue).

In summary, we propose the DNZ acoustic-cloaking cells for one-dimensional invisibility that perfectly eliminate the perceptibility of the hidden objects from underwater sounds. The plane-wave feature is maintained throughout without distortion and reflection is completely suppressed due to EST. The DNZ property at systematic resonances is generally explained by the inherent combination of the acoustic inertance and the acoustic compliance of the structure. Note that the DNZ cell for cloaking is built only by a simple layout with a uniform material, while its performance is independent of arbitrary objects (location, shape, etc.)

inside it. Moreover, by placing additional DNZ cells, we can expand and design the overall cloaked space without upper limit and with an arbitrary distribution, in free space as well as in waveguides. The mechanical interpretation: the spring-mass oscillation is also employed to further discuss the geometry-dependent resonances. We believe the proposed DNZ cell enables a distinct and concise way of realizing robust and stable acoustic cloaking by viable bulky materials.

**References.**

[1] J. B. Pendry, D. Schurig, and D. R. Smith, *Science* **312**, 1780 (2006).

[2] U. Leonhardt, *Science* **312**, 1777 (2006).

[3] D. Schurig, J. J. Mock, B. J. Justice, S. A. Cummer, et al. *Science* **314**, 977 (2006).

[4] S. Xu, X. Cheng, S. Xi, R. Zhang, et al. *Phys. Rev. Lett.* **109**, 223903 (2012).

[5] S. A. Cummer, B. I. Popa, D. Schurig, D. R. Smith, et al. *Phys. Rev. Lett.* **100**, 024301 (2008).

[6] S. Zhang, C. Xia, and N. Fang, *Phys. Rev. Lett.* **106**, 024301 (2011).

[7] B. I. Popa, L. Zigoneanu, and S. A. Cummer, *Phys. Rev. Lett.* **106**, 253901 (2011).

[8] V. M. García-Chocano, L. Sanchis, A. Díaz-Rubio, J. Martínez-Pastor, et al. *Appl. Phys. Lett.* **99**, 074102 (2011).

[9] L. Sanchis, V. M. García-Chocano, R. Llopis-Pontiveros, A. Climente, et al. *Phys. Rev. Lett.* **110**, 124301 (2013).

[10] J. J. Park, K. J. B. Lee, O. B. Wright, M. K. Jung, et al. *Phys. Rev. Lett.* **110**, 244302 (2013).

[11] R. Fleury, and A. Alù, *Phys. Rev. Lett.* **111**, 055501 (2013).

[12] R.V. Craster, and S. Guenneau (Ed.), *Acoustic Metamaterials* (Springer, 2013).

[13] S. Tachi, Telexistence and Retro-reflective Projection Technology, Proceedings of the 5[th] Virtual Reality International Conference, Laval Virtual, France, May 13-18, 2003.

[14] D. T. Blackstock, *Fundamentals of Physical Acoustics* (Wiley, New York, 2000).

[15] S. H. Lee, C. M. Park, Y. M. Seo, Z. G. Wang, et al. *Phys. Rev. Lett.* **104**, 054301 (2010).

[16] R. Fleury, D. L. Sounas, C. F. Sieck, M. R. Haberman, and A. Alù, *Science* **343**, 516 (2014).

[17] X. Wang, *Appl. Phys. Lett.* **96**, 134104 (2010).

[18] L. E. Kinsler, A. U. Frey, A. B. Coppens, and J. V. Sanders, *Fundamental of Acoustics* (Wiley, New York, 1982).

[19] Z. Yang, J. Mei, M. Yang, N. H. Chan, and P. Sheng, *Phys. Rev. Lett.* **101**, 204301 (2008).